\documentstyle[12pt]{article}

\newcommand{\la}[1]{\label{#1}}
\newcommand{\beq}{\begin{equation}}
\newcommand{\eeq}{\end{equation}}
\newcommand{\ba}{\begin{eqnarray}}
\newcommand{\ea}{\end{eqnarray}}
\newcommand{\bea}{\begin{eqnarray*}}
\newcommand{\eea}{\end{eqnarray*}}

\newcommand{\eq}{eq.~}

\newcommand{\fig}{fig.~}

\newcommand{\simls}{\hbox{$\,$\raise.5ex\hbox{$<$}
       \kern-1.1em \lower.5ex\hbox{$\sim$}$\,$}}
\newcommand{\simgt}{\hbox{$\,$\raise.5ex\hbox{$>$}
       \kern-1.1em \lower.5ex\hbox{$\sim$}$\,$}}
\let\oref=\ref
\renewcommand{\ref}[1]{(\oref{#1})}
\newcommand{\del}{\partial}

\makeatletter
\def\fnum@figure{Fig.~\thefigure}
\newdimen\ctcleftskip \ctcleftskip=.4em
\newdimen\ctcd@pth \newdimen\ctcsk@p \ctcsk@p=0pt
\def\c@ntract#1#2{\ctcd@pth#2\relax\mathord{\vtop{\ialign{##\crcr
$\displaystyle{#1}$\crcr\noalign{\kern1pt\nointerlineskip}
\ctcf@ll\crcr}}}}
\def\ctcf@ll{\hskip\ctcsk@p\hskip\ctcleftskip\vrule\@height\ctcd@pth
\@depth\z@\leaders\hrule\hfill%
\vrule\@height\ctcd@pth\@depth\z@\hskip\ctcleftskip\relax}
\newdimen\ctcnormaldepth \ctcnormaldepth=2.5pt
\def\contract#1{\c@ntract{#1}{\ctcnormaldepth}}
\def\xcontract#1#2{\c@ntract{#2}{#1\ctcnormaldepth}}
\def\doublecontract#1#2#3{
\setbox\z@\hbox{$\displaystyle\contract{#1#2}$}
\setbox\@ne\hbox{$\displaystyle#1$}\dp\z@\dp\@ne\ctcsk@p=\wd1
\setbox\@ne\hbox{$\displaystyle\xcontract2{\box\z@#3}$}
\box1\ctcsk@p\z@}
\def\innercontract#1{
\setbox\z@\hbox{$\displaystyle\contract{#1}$}
\setbox\@ne\hbox{$\displaystyle #1$}\dp0=\dp1\box\z@}
\makeatother

\begin{document}
\pagestyle{empty}
\setlength{\parindent}{0.0cm}

% \begin{minipage}{13cm}
\begin{center}
\vspace*{2.2cm}
{\Large\bf The 2--Loop Effective  Potential of} \\
\vspace*{2mm}
{\Large\bf the 3d SU(2)--Higgs Model} \\
\vspace*{2mm}
{\Large\bf in a General Covariant Gauge} \\
\vspace*{12mm}
{\large\bf M.~Laine\footnote{
Electronic address: mlaine@phcu.helsinki.fi}} \\
\vspace*{3mm}
{\sl  Department of Theoretical Physics, \\
P.O.~Box 9, FIN-00014 University of Helsinki, Finland} \\
\vspace*{3mm}
9 June 1994
\end{center}

\vspace*{-9.8cm}
\hfill Preprint HU-TFT-94-20
\vspace*{9.5cm}

\begin{center}
{\large\bf Abstract}
\end{center}
\vspace*{2mm}
To study the convergence of the loop expansion at the high--temperature
electroweak phase transition, we calculate the 2--loop
effective potential of the 3d SU(2)--Higgs model in a general
covariant gauge. We find that the loop expansion
definitely breaks down for large $\xi$, but converges rather
well for smaller values, deep in the broken phase.

%PACS numbers: 98.80.Cq
%\end{minipage}

\newpage
\pagestyle{plain}
\setcounter{page}{1}
\setcounter{footnote}{0}
\setlength{\parindent}{0.85cm}

Recently, both perturbative~\cite{pert,jakovac1,fkrs1} and
non--perturbative~\cite{nonpert,fkrs2} studies of the high--T electroweak
phase transition have been performed, with the indication
that perturbation theory works poorly in the symmetric
phase, due to non--perturbative effects. The purpose of this note
is to investigate the convergence of loop expansion in the
broken phase. The method is to calculate the high--T
asymptotic of the 2--loop effective potential
in a general covariant gauge, and to extract from it a
quantity which is gauge--independent in the full theory.
The degree of gauge--dependence at the 2--loop level is then
expected to tell something about the convergence of the expansion.
In accordance with this philosophy, we do not convert the loop
expansion into an expansion in the coupling constants at any stage.
The convergence of loop expansion in the broken phase
has also been studied in ref.~\cite{fkrs2},
by comparing lattice data and perturbation theory.

Our starting point is the action
\beq
S=\int d^3x \left[\frac{1}{4}F^a_{ij}F^a_{ij}+
(D_i\Phi)^{\dagger}(D_i\Phi)+[m^2(\mu)+\delta m^2]\Phi^{\dagger}\Phi+
\lambda(\Phi^{\dagger}\Phi)^2 \right]
\la{action}
\eeq
with $F^a_{ij}=\del_iA_j^a-\del_jA_i^a+g\epsilon^{abc}A^b_iA^c_j$
and $D_i\Phi=(\del_i-ig\tau^aA^a_i/2)\Phi$.
Here the $\tau^a$:s are the Pauli matrices. The only $\Phi$--dependent
infinity in this theory is removed by $\delta m^2$, and
the coupling constants are RG--invariant. Gauge fixing
and compensation is achieved by adding to \eq\ref{action} the
term
\beq
S_g=\int d^3x\left[\frac{1}{2\xi}(\del_iA_i^a)^2+
\del_i\bar{c}^a\del_ic^a+g\epsilon^{abc}\del_i\bar{c}^aA_i^bc^c
\right] \,\, .
\la{gf}
\eeq
All the fields have dimension
$[\mbox{GeV}]^{1/2}$,
and $\lambda$ and $g^2$ have dimension $[\mbox{GeV}]$.

The action $S+S_g$ is an effective field theory
in the sense of ref.~\cite{weinberg},
obtained from the gauge--Higgs sector of the standard
model after integrating out the heaviest degrees of freedom,
namely, the non--zero Matsubara frequencies and the $A_0^a$
field\footnote{We calculated $V(\varphi)$ also in the theory from
which $A_0^a$ has not been integrated out, but for simplicity
we discuss only eq.~\ref{action} in detail. The differences between
the results are addressed below.}~\cite{jakovac1, fkrs1}.
The relations between
the parameters of our 3d theory and those of the Standard Model
are given in ref.~\cite{fkrs1}.
To the assumed accuracy, these relations are
gauge--invariant~\cite{fkrs1,jakovac2}.
The temperature dependence of
coupling constants is $\lambda,\, g^2\propto T$,
and that of $m^2(\mu)$ is
\beq
m^2(\mu)=\gamma(T^2-T_0^2)+\frac{1}{16\pi^2}f_{2m}\log
\left(\frac{3T}{\mu}\right) \,\, .
\la{msq}
\eeq
Here $f_{2m}$ is the coefficient of
the 2--loop mass--counterterm, proportional
to $T^2$ when expressed in terms of 4d couplings. Notice that
the gauge fixing parameter $\xi$ gets renormalized in the
reduction step.

To calculate the effective potential $V(\varphi)$, one writes
$\Phi=[\phi_3+i\phi_4, \varphi+\phi_1+i\phi_2]^T/\sqrt{2}$
in the action $S+S_g$
and neglects terms linear in quantum fields~\cite{jackiw}.
This defines a new theory with the masses
$m_1^2\equiv m^2(\mu)+3\lambda\varphi^2$,
$m_2^2\equiv m^2(\mu)+\lambda\varphi^2$, and
$m_T^2\equiv g^2\varphi^2/4$. The non--trivial propagators are
\ba
\contract{{A_i^a(-k)}{A_j^b}}(k) & = & \delta^{ab}\left[
\frac{\delta_{ij}-k_ik_j/k^2}{k^2+m_T^2}+
\xi\frac{k_ik_j}{k^2}\frac{k^2+m_2^2}{k^2(k^2+m_2^2)+
\xi m_T^2m_2^2}\right] \nonumber \\
\contract{{\phi_G(-k)}{\phi_G}}(k) & = &
\frac{k^2+\xi m_T^2}{k^2(k^2+m_2^2)+
\xi m_T^2m_2^2}\,\, , \,\,\,\,\,\,\,\, G=2,3,4 \la{propagators} \\
\contract{{\phi_2(k)}{A^3_i}}(-k) & = &
-\contract{{\phi_3(k)}{A^2_i}}(-k) =
-\contract{{\phi_4(k)}{A^1_i}}(-k) = \frac{i\xi m_Tk_i}{k^2(k^2+m_2^2)+
\xi m_T^2m_2^2} \,\, .\nonumber
\ea
It is seen that for positive $\xi$,
$m_2^2$ must be positive. Hence the calculation is valid only
for larger values of $\varphi$ than the tree--level minimum,
where $m_2^2=0$.

At this point,
our choice of gauge deserves some explanation. In the literature,
one sometimes uses an $R_{\xi}$--type of a gauge,
since then the non--diagonal terms between the longitudinal parts of
the $A_i^a$--fields and the Goldstone bosons are cancelled,
yielding much simpler propagators. This means that one chooses
a different gauge for each different value of $\varphi$ in the effective
potential. It is not {\em a priori} clear that this is allowed (for more
criticism, see ref.~\cite{arnold}). However, at least in the case
of scalar electrodynamics, it can be proved
that the error induced is of higher order in $\hbar$~\cite{kugo}
(for a more pessimistic conclusion, see ref.~\cite{nielsen}).
To be on the safe side, we chose the $(1/2\xi)(\del _iA_i^a)^2$--gauge
for our non--Abelian calculation.

The 2--loop graphs to be calculated are shown in \fig1.
Due to the fact that the $\phi\phi\phi$-- and
$\phi AA$--vertices include at least one Higgs field $\phi_1$,
and only the Goldstone fields can transform into longitudinal
vector bosons, four possible graphs (which are not shown)
vanish. Also, the graph (f2.a) vanishes due to antisymmetry
in momentum integration. The method of calculation for
the remaining graphs is to write
\beq
\frac{1}{k^2(k^2+m_2^2)+\xi m_T^2m_2^2}=
\frac{1}{m_2^2(R_+^2-R_-^2)}\left[\frac{1}{k^2+m_2^2R_-^2}-
\frac{1}{k^2+m_2^2R_+^2}\right]
\,\, , \la{method}
\eeq
where $R^2_{\pm}=1/2\pm\sqrt{1/4-\xi(m_T/m_2)^2}$,
and then to use standard Landau--gauge values of integrals,
as presented e.g.~in ref.~\cite[App.B.2]{fkrs1}
(we use dimensional regularisation).
Note that $R^2_{\pm}$ can be complex, but this does not
matter, since the poles are off the integration path and
the final result is real. The 1--loop effective potential is
\beq
V(\varphi)=\frac{1}{2}m^2(\mu)\varphi^2+
\frac{1}{4}\lambda\varphi^4-
\frac{1}{12\pi}[6m_T^3+m_1^3+3m_2^3(R_+^3+R_-^3)]
\,\, . \la{1loop}
\eeq
Due to lack of space, the 2--loop part is presented
explicitely elsewhere (ref.~\cite{ml}).
%The 2--loop potential is presented in the Appendix.

There are two basic ways to check the 2--loop result. First,
in the limit $\xi\to 0$ it reproduces the Landau--gauge result
as given in ref.~\cite{fkrs1}. Second, due to the facts that
we are using a gauge--invariant regularization procedure, and
the term $\Phi^{\dagger}\Phi$ is gauge--invariant, the
counterterm $\delta m^2$ must be gauge--independent. Indeed,
the terms proportional to $g^4\xi$ and $g^4\xi^2$ from
graphs (a), (f1), (d1), (x3), (x4) and (x5) cancel, and the
divergent piece is
\beq
\frac{1}{16\pi^2}\frac{\mu^{-4\epsilon}}{4\epsilon}\left\{
\frac{\varphi^2}{2}\left[
\frac{51}{16}g^4+9\lambda g^2-12\lambda^2\right]+
3g^2m^2(\mu)\right\}
\,\, . \la{counterterm}
\eeq
The quantity inside the square brackets
is just $f_{2m}$ appearing in \eq\ref{msq}.

As was noted above, the integrals appearing in the calculation of
$V(\varphi)$ are defined only for $m_2^2>0$, if $\xi\neq0$.
Since we are interested in making calculations even below
the temperature where the symmetric phase is metastable, and
in our 3d theory this roughly means that $m^2(\mu)<0$,
there is a region of $\varphi$
where $m_2^2=m^2(\mu)+\lambda\varphi^2\to0$.
It is seen from \eq\ref{propagators} that in this limit integrals
including Goldstone bosons become IR--divergent, with a
gauge--dependent coefficient. In the effective potential,
this causes part of the diagrams to diverge, as $m_2^2\to0$.
The most severe divergences, proportional to
$g^2(m_T^3/m_2)\xi^{5/2}$,
come from diagrams (f1) and (x4), but these cancel. Logarithmic
divergences proportional to $g^2m_T^2\xi^2\log (m_2)$, coming from
the same diagrams, also cancel, as do all the divergences proportional
to $\lambda(m_T/m_2)^{1/2}(m_T^3/m_1)\xi^{7/4}$, coming from diagrams
(c), (f1), and (x2). Still, a divergent piece
\beq
\frac{1}{16\pi^2}\left[\frac{9}{8}
\lambda m_T^2\xi^{3/2}\left(\frac{m_T}{m_2}\right)-
\frac{9}{4\sqrt{2}}g^2m_T^2\xi^{3/4}\left(\frac{m_T}{m_2}\right)^{1/2}
\left(1+4\sqrt{2}\frac{\lambda^{3/2}}{g^3}\right)\right]
\la{divergent}
\eeq
from diagrams (h4), (f2), and (c)
remains. Even the 1--loop potential is non--analytic at the
$m_2^2\to0$ limit, since its $m_2$--dependence is of the form
$\xi^{3/4}m_2^{3/2}m_T^{3/2}$. The divergence in \eq\ref{divergent}
is gauge--dependent, and would
not appear at all e.g.~in the $R_{\xi}$--gauge.

Let us briefly note that in ref.~\cite{bfh}, a gauge--invariant
effective potential of the composite operator $\Phi^{\dagger}\Phi$
is calculated at 1--loop level using the saddle point method.
One of the two saddle points is at $m_2^2=0$.
A 2--loop calculation of $W[J]$ has exactly the same Feynman rules
and propagators as our calculation, but the set of diagrams is
different, since in their calculation all the connected vacuum
diagrams are to be included, whereas only one--particle--irreducible
diagrams enter our calculation~\cite{jackiw}. If their 2--loop
calculation is to be successful in the covariant gauge,
the divergent piece of \eq\ref{divergent}
must therefore be cancelled by the extra diagrams.

For the present calculation, the crucial question is whether the
IR-divergences in \eq\ref{divergent} completely spoil the
perturbation theory even for $m_2^2>0$. This cannot be
quite so since the $\xi\to0$--limit is the well-behaved Landau
gauge. From the formulas for $R^2_{\pm}$, it is seen that one is
effectively at this limit when
$\xi\ll(m_2/2m_T)^2\simls\lambda/g^2$.
This would indicate that
only extremely small values of $\xi$ are allowed, since
typically at the broken minimum $m_2=(0.2-0.4)m_T$.
However, due to the cancellations, larger values are also possible.
The remaining $1/m_2$--divergence in \eq\ref{divergent}
is proportional to $\lambda$, and its effect is damped even more
by the fact that the $1/m_2^{1/2}$--divergences come with the
opposite sign. The dangerous term is
$-({1}/{16\pi^2})(9g^2m_T^{5/2}\xi^{3/4})/{4(2m_2)^{1/2}}$;
comparing it with the most important gauge--independent
2--loop terms, which are proportional to $g^2m_T^2$, we see that it
is sufficient to have $\xi\simls(m_2/m_T)^{2/3}$. The finite
terms further suppress the divergent pieces at $m_2^2\neq0$,
so that numerically their effect is confined to a very
narrow region near $m_2^2=0$, as is seen from \fig2.

Increasing $\xi$ still further, the loop expansion definitely breaks down.
Namely, by just plotting the effective
potential, one can see that the location of the broken
minimum, $\langle\varphi\rangle$, gets smaller for larger $\xi$.
For very large $\xi$, this minimum gets so near the point
$m_2^2=0$ that it is in the region where the divergences are dominant.
The 1--loop potential ceases to have any minimum at all
in the region $m_2^2>0$! To estimate the value of $\xi$ for
which this happens, one can write all the $\varphi$-dependence in
the equation $dV_{\rm 1-loop}(\varphi,\xi\gg\lambda/g^2)/d\varphi=0$ in
terms of $m^2_2$ to get (notice that $m^2<0$)
\beq
\xi^{3/4}g^{3/2}\lambda^{1/4}\frac{2m_2^2-m^2}
{m_2^{1/2}(m_2^2-m^2)^{1/4}}=
\frac{g^3}{\lambda^{1/2}}(m_2^2-m^2)^{1/2}+
4\lambda(3m_2^2-2m^2)^{1/2}-\frac{16\pi}{3}m_2^2 \,\, .
\la{nominimum}
\eeq
By inspecting the functional form of the LHS and RHS
of \eq\ref{nominimum} one sees that for large $\xi$, there
is no solution with $m_2>0$ (For smaller $\xi$, there are
two solutions; the smaller of these is unphysical, see \fig2).
To get a numerical value, one can expand \eq\ref{nominimum} to
order $m_2^2$, neglect even the $m_2^{3/2}$--term for an upper
bound on $\xi$, assume that $\lambda\ll g^2$, and go somewhat
below the temperature $T_0$. This leads to
\beq
\xi_{\rm max}\approx(\hat{g}^3/\hat{\lambda}^{7/6})
(T/200\sqrt{-m^2})^{1/3}
\,\, , \la{ximax}
\eeq
with coupling constants scaled
dimensionless by $T$. This means that for small Higgs masses
and for temperatures near the point where the symmetric
phase becomes metastable, $\xi$ can be quite large.

Having estimated the range of validity of the 2--loop
potential, we now study the convergence of the loop expansion.
The object which hopefully converges well is not $V(\varphi,T,\mu)$
itself, but a renormalization group improved version of it. This
means that one calculates $\del V(\varphi,T,\mu)/\del\varphi$,
chooses $\mu$ for each $\varphi$ so that large logarithms are
killed, and then integrates to get an
improved $V(\varphi,T)$~\cite{fkrs1,drtj}. In practice,
one takes $\mu\propto m_T$.
One could kill the large logarithms directly in
the potential $V(\varphi,T,\mu)$, but in this case
what used to be a $\mu$--dependent vacuum part,
becomes $\varphi$--dependent, and has
to be taken into account~\cite{drtj}.

Neither the effective potential at any temperature, nor
its stationary points, are gauge--independent
quantities~\cite{jackiw,kugo,nielsen}.
However, the value of the effective potential
at a stationary point is gauge--independent. This value is related
to another gauge--independent quantity,
$\langle\Phi^{\dagger}\Phi\rangle$. This is the quantity which we
choose as our indicator of the convergence of perturbation theory:
if we find at 2--loop level that $\langle\Phi^{\dagger}\Phi\rangle$
depends only weakly on the gauge parameter, we have reason
to believe that the loop expansion converges rather well.

For bare quantities, one gets the equation
$\langle\Phi^{\dagger}\Phi\rangle_B =
{dV_B(\varphi_{\rm min}(T),T)}/{dm_B^2}$.
Unfortunately, renormalization in a sense
spoils this relation, since the $\varphi$--independent but
$m^2(\mu)$--dependent counterterm in \eq\ref{counterterm} has to be
removed. This also defines
the renormalization of $\langle\Phi^{\dagger}\Phi\rangle_B$. As a result,
$\langle\Phi^{\dagger}\Phi\rangle$ becomes $\mu$--dependent, and thus
physically ill--defined in the continuum theory.
However, it can be related to a corresponding quantity on the
lattice [ref.~\cite{fkrs2}, eq.~(13)], and thus given a physical
meaning. Different values of $\mu$ just correspond
to different lattices. To relate this $\mu$--dependence
to the use of RG--improved
perturbation theory, notice that the location of the minimum
of $V(\varphi)$ is $\mu$--independent (though $\xi$--dependent)
in the exact theory. Therefore, we use the
RG--improved potential to find the best possible approximation
for the location of the minimum, and then plug this value
into the quantity $dV(\varphi,T,\mu)/dm^2(\mu)$ with a fixed $\mu$,
to get $\langle\Phi^{\dagger}\Phi(\mu)\rangle$. In practice, one
cannot choose $\mu$ quite arbitrarily, since we are working at a finite
order in the loop expansion.

In \fig3, the quantities $\langle\varphi\rangle$
and $\sqrt{2\langle\Phi^{\dagger}\Phi\rangle}$ for $m_H=80$ GeV
are shown at different
temperatures below $T_c$. One sees
that $\sqrt{2\langle\Phi^{\dagger}\Phi\rangle}$
is very accurately gauge--independent. We have also checked that
there is less gauge-dependence at 2--loop level than at 1--loop level.
One can think of two reasons for the remaining gauge--dependence.
First, by comparing the most important terms in the
effective potential, one sees that the effective loop--expansion
parameter in the Landau gauge is $\hat{g}^2T/m_T(\varphi)$. Since
at high temperatures $\langle\varphi\rangle$ gets smaller and
$T$ gets larger, the loop expansion converges
worse, and there is more gauge dependence, as is observed
in \fig3. Second, the rather strong
gauge dependence at large $\xi$ is caused by the fact that one
is approaching the point where the loop expansion breaks down
due to IR--divergences.

In \fig3, $\sqrt{2\langle\Phi^{\dagger}\Phi\rangle}$ as calculated from
the theory with $A_0^a$ included, is also shown. The allowed range of
$\xi$ is larger, as one can see from the same argument
which led to \eq\ref{ximax}. There is a bit
more gauge dependence in these curves, indicating that
an error has been induced in
constructing the $A_0^a$--integrated--out theory.

Varying the Higgs mass, one can see that for smaller Higgs masses
the allowed range of $\xi$ is much larger, as predicted by
\eq\ref{ximax}. The absolute value of the
derivative $d\langle\Phi^{\dagger}\Phi\rangle/d\xi$
is slightly larger for, say, $m_H=35$ GeV, than for $m_H=80$ GeV.
% This is probably due to the fact that when
% $m_H\neq m_W$, our choice of taking $\mu\propto m_T$ in
% RG--improvement is not enough to kill all the large logarithms.

Although it is not expected that the loop expansion
converges very well in the symmetric phase,
we have drawn\footnote{At 1--loop level, a similar
study was made in ref.~\cite{jakovac2}. See also ref.~\cite{arnold}.}
in \fig4 the critical temperature $T_c$ and
$\langle\varphi_c\rangle/T_c$ as a function of $\xi$.
The gauge dependence is clearly stronger than for
$\sqrt{2\langle\Phi^{\dagger}\Phi\rangle}$ deep in the broken phase.
For large $\xi$ the loop expansion again breaks down. The reason
is that at large $\xi$,
$T_c$ is actually below the temperature
$T_0$ in \eq\ref{msq}. The logarithmic term makes
$m^2(\mu)$ positive for very small $\varphi$, since $f_{2m}$
is positive and $\mu$ is proportional to $\varphi$.
When $\varphi$ grows larger, $m^2(\mu)$ becomes
negative, but the term $\lambda\varphi^2$
in $m_2^2$ grows larger. The result is that for a large
range of $\varphi$, $m_2^2$ is very close to zero. Hence,
the loop expansion is unreliable due to IR--divergences.

To conclude, our results
indicate that the loop expansion
breaks down for too large $\xi$, but converges
relatively well deep in the broken phase for smaller values.
For more definite conclusions, one must compare
perturbative results with lattice calculations~\cite{fkrs2}.
Good convergence of the loop expansion,
absence of IR--divergences, and
sheer calculational simplicity, strongly support the
use of Landau gauge.

%\section*{Acknowledgements}
{\bf Acknowledgements.} The motivation for this work, as well as
invaluable support during its completion, was provided by
M.~Shaposhnikov, for which I thank him. I am also grateful to
K.~Kajantie for discussions, and to CERN for hospitality.

\end{document}